# Droplet impact and splitting behaviour on superhydrophobic wedges

**Gudlavalleti V V S Vara Prasad** [1,3], **Parmod Kumar** [3, *, c], **Purbarun Dhar** [2, *, b] and **Devranjan Samanta** [1,*, a]

[1] Department of Mechanical Engineering, Indian Institute of Technology Ropar, Punjab–140001, India

[2] Hydrodynamics and Thermal Multiphysics Lab (HTML), Department of Mechanical Engineering, Indian Institute of Technology Kharagpur, West Bengal–721302, India

[3] School of Mechanical and Materials Engineering, Indian Institute of Technology Mandi, Mandi, Himachal Pradesh–175075, India

*Corresponding authors:

[a]E-mail: devranjan.samanta@iitrpr.ac.in, [a]Tel: +91-1881-24-2109
[b]E-mail: purbarun@mech.iitkgp.ac.in, [b]Tel: +91-3222-28-2938
[c]E-mail: parmod@iitmandi.ac.in, [c]Tel: +91-1905-26-7858

**Abstract**

We report an extensive computational and experimental investigation of droplet impact and subsequent splitting hydrodynamics on superhydrophobic (SH) wedges. 2D and necessary 3D simulations using the volume-of-fluid (VOF) method, backed with experimentations, have been performed to predict the droplet impact, spreading, split-up, retraction against sliding, and daughter droplet lift-off events from the SH wedge. In particular, we examine how the wedge angle ($\phi$), wedge asymmetry ($\phi_1 - \phi_2$), Weber number ($We$) and normalized Bond number ($Bo^*$) influence the post-impact dynamics. We observe that for symmetric wedges, the maximum spread factor ($\beta_{max}$) of the droplet decreases with an increase in wedge angle ($\phi$) at a fixed $We$. At high wedge angles, the sharp steepness of the wedge causes less contact area for the droplet to spread. For the asymmetric wedges, it has been noted that $\beta_{max}$ increases with an increase in the $We$ due to the higher inertial forces of the droplet against sliding. Furthermore, the $\beta_{max}$ increases with an increase in $Bo^*$ at a fixed $We$ due to the dominance of the gravitational force over the capillary force of the droplet. It has been also found that at the same $Bo^*$, the $\beta_{max}$ increases with an increase in $We$ due to the dominance of inertial forces over the capillary forces. The split volume of daughter droplets during the split-up stage for different symmetric and asymmetric wedge angles has been discussed. In general, our 2D simulations agree well with the experiments for a major part of the droplet's lifetime. Further, we have conducted a detailed 3D simulation-based energy budget analysis to estimate the temporal evolution of the various energy components at different post-impact hydrodynamic regimes.

# 1. Introduction

Droplet impact on solid surfaces [1] has broad implications in a wide variety of applications; viz. inkjet printing [2]–[6], spray coating and painting [7], spray cooling [8]–[11], anti-icing [12], etc. Consequently, studies focusing on the allied transport phenomena of droplet impact and its outcomes; involving experiments, theory, and computer simulations have been widely reported in literature. Given the large range of physical insights available from detailed simulations[13–16], several research teams have made significant contributions to the underlying physics of droplet spreading, deformation, fragmentation, rebound, and recoil on solid surfaces of different wettabilities.

Hu et al. [17] simulated the impact of microdroplets on structured superhydrophobic (SH) surfaces using the Volume of fluid (VOF) method and showed that microdroplets exhibited higher Laplace pressure than macro-droplets, resulting in an increased wetting pressure during the impalement event. Debnath et al. [18], [19] investigated the spreading dynamics of highly viscous droplets on SH surfaces using the VOF method. They reported the dominant influence of viscosity on the spreading of such droplets, with minimal effects of inertia. In another study, they [20] examined the effect of surface inclination using the VOF method and noted a significant decrease in the droplet contact time with an increase in the inclination angle. Henman et al. [21] explored the early transients of droplet impact dynamics on textured and lubricant-infused surfaces, employing the VOF method, and showed that increasing the spacing between surface asperities led to a reduction in the horizontal range of the splash jet.

Vontas et al. [22] conducted a VOF study of droplet impact on smooth surfaces with varying wettability. They showed the predictive capability of Kistler's dynamic contact model towards capturing droplet spreading, recoiling, and rebounding during the post-impact phase. Malgarinos et al. [23] reviewed the contact angle dynamics during droplet spreading, focusing on different models ranging from standard to non-wetting models. Their comparison showed consistent results, except for Shikhmurzaev's dynamic contact angle model, which frequently yielded more accurate spreading ratio predictions. Zhang et al. [24] reported multi-scale simulations of dynamic wetting by combining Molecular Dynamics (MD) and VOF approaches. They showed that for simulation of the wetting behavior of nanodroplets on surfaces, information of the dynamic contact angles alone was insufficient to capture molecular-level effects.

Liu et al. [25], [26] investigated droplet impact dynamics on spherical and concave surfaces (VOF method). Their findings exhibited good agreement with experimental results. On concave surfaces they observed a minor initial increase in the maximum spreading factor and area, followed by a decrease, as the target diameter decreased. Ding et al. [27] simulated (VOF) droplet breakup and rebound dynamics during impact on small cylindrical SH targets. Their work indicated that higher $We$ accelerated droplet spreading and promoted breakup. Luo et al. [28] simulated (VOF) droplet impact dynamics on a cone and observed a ~54% reduction in contact time during the phase where the droplet lifted from the surface as a ring, compared to the impact on a flat surface.

Wasserfall et al. [29] numerically studied the coalescence-induced droplet jumping on a SH surface and noticed that reducing the droplet size ratio enhances the ratio of adhesion forces between the merged droplets to the excess energy released during the merging process. Russo et al. [30] numerically examined the droplet impact dynamics on wettability-patterned surfaces by implementing Kistler contact angle model, and showed a good match of the simulation with experiments on droplet translation, splitting, and vectoring (nonorthogonal rebound of droplets impacting orthogonally on a SH domain). Wang et al. [31] simulated the droplet impact on a mesh array and investigated the effect of liquid properties on impact, penetration, and fragmentation using the Lattice Boltzmann Method (LBM). Hao et al. [32] examined the droplet splashing dynamics on an inclined surface and noted that splashing can be entirely suppressed either by increasing the inclination angle or by reducing the ambient pressure. Zhao et al. [33] numerically studied the effect of the impact velocity on spreading area and droplet kinetics of oblique droplet impact on a horizontal solid surface. They also highlighted the influence of tangential velocity on viscous dissipation and showed that large tangential velocity results in increased viscous dissipation.

Despite numerous investigations on droplet impact on various SH surfaces like flat, spherical, cylindrical, and conical ones[25–27], [34], the impact hydrodynamics of droplets on SH wedges have been relatively unexplored. We pose an interesting question here; as to why the simulation and experiments of droplet impact dynamics on a SH wedge are scientifically and academically important? On a SH wedge, the post-impact dynamics are markedly different from the impact configurations tested thus far in the literature. This stems from the fact the post-impact events eventually and mandatorily lead to the splitting of the droplet into daughter droplets. This split could be symmetric for symmetric wedges, and asymmetric for wedges having different base angles; with the event of splitting strongly dependent on the wedge asymmetry. The point of scientific interest lies on the comprehension of the fluid dynamics of this splitting paradigm, especially in the case of asymmetric wedges. The role of the geometry, velocity and pressure conditions that lead to the nature and evolution of the split may add important mechanics to the rich physics of droplets and their interactions with obstacles.

In this research, we have simulated the drop impact dynamics on SH wedges using the VOF method. We have also conducted detailed experiments using high-speed imaging and validated the accuracy of the simulations against experimental data. To replicate the impact dynamics with greater accuracy, we have performed 3D simulations, as and wherever essential, to establish the events post-impact, during and after splitting. A thorough investigation has been conducted to explore the effects of the wedge angle $(\phi)$, the wedge base angles $(\phi_1, \phi_2)$ and their asymmetry $(\phi_1 - \phi_2)$, impact Weber number ($We$), and normalized Bond number ($Bo^*$) on the drop impact dynamics. The energy budget estimation during post-impact and splitting hydrodynamics of the droplet has been incorporated also to explain the different outcomes.

## 2. Simulation methodology
### 2.1. Governing equations

We have simulated the interaction between the droplet (water)-gas (air) pair during the process of droplet impact on a stationary, SH wedge. The simulation of the post-impact and splitting hydrodynamics of the droplet is dealt with in two different phases. As the interface separates both phases, it is essential to choose an appropriate interface-tracking method. Despite the various possible numerical techniques, such as the two-fluid method, level-set method and VOF method, the VOF method (Hirt and Nichols [13]) has been adopted due to its higher accuracy in capturing interface evolution and deformation. The governing equations to be solved are as follows:

$$\nabla \cdot \vec{V} = 0 \tag{1}$$

$$\frac{\partial \alpha}{\partial t} + \nabla \cdot (\alpha \vec{V}) = 0 \tag{2}$$

$$\rho \left[\frac{D\vec{V}}{Dt}\right] = -\nabla p + \nabla \cdot \mu \left[(\nabla \vec{V}) + (\nabla \vec{V})^T\right] + \rho \vec{g} + \vec{F}_{st} \tag{3}$$

Where, $\vec{V}, p, \vec{g}, \vec{F}_{st}, \alpha$ represents the velocity, pressure, gravitational acceleration, volumetric surface tension force, and phase volume fraction, respectively.

In this context, the liquid volume fraction $(\alpha_l)$ and the corresponding gas volume fraction $(\alpha_g)$ within each control volume are expressed as follows:

$$\alpha_i = \begin{cases} 0 & if\ the\ cell\ is\ occupied\ by\ only\ gas(air) \\ 0 < \alpha_i < 1 & if\ the\ cell\ contains\ both\ gas(air)\ and\ liquid\ (water) \\ 1 & if\ the\ cell\ is\ occupied\ by\ only\ liquid(water) \end{cases} \tag{4}$$

and

$$\alpha_l + \alpha_g = 1 \tag{5}$$

The physical properties of the mixture are determined as a weighted average based on $\alpha$. The density and viscosity of the equivalent fluid are obtained as:

$$\rho = \alpha_l \rho_l + (1 - \alpha_l)\rho_g \tag{6}$$

$$\mu = \alpha_l \mu_l + (1 - \alpha_l)\mu_g \tag{7}$$

where, $\rho_l$, $\mu_l$, are the density and viscosity of the liquid phase, and $\rho_g$, $\mu_g$ are the density and viscosity of the gaseous phases, respectively.

The VOF method incorporates a customized interface compression scheme, wherein a supplementary convective term is introduced to the transport equation of the volume fraction as follows:

$$\frac{\partial \alpha}{\partial t} + \nabla \cdot (\alpha \vec{V}_l) = 0 \tag{8}$$

$$\frac{\partial (1-\alpha)}{\partial t} + \nabla \cdot \left((1-\alpha)\vec{V}_g\right) = 0 \tag{9}$$

$$\vec{V} = \alpha \vec{V}_l + (1 - \alpha)\vec{V}_g \tag{10}$$

The velocity of the equivalent fluid is determined by assuming it to be the weighted average based on the volume fraction as

$$\frac{\partial \alpha}{\partial t} + \nabla \cdot (\alpha \vec{V}) + \nabla \cdot \left(\alpha(1-\alpha)(\vec{V}_l - \vec{V}_g)\right) = 0 \tag{11}$$

In the equation (3), the volumetric surface tension force $F_{st}$ acting at the liquid-gas interface is obtained from the continuum surface (CSF) model (Brackbill et al. [14]) as

$$\vec{F}_{st} = \gamma k \nabla \alpha \tag{12}$$

$$k = -\nabla \cdot \left(\frac{\nabla \alpha}{|\nabla \alpha|}\right) \tag{13}$$

Where $k$ is the mean curvature of the interface. The final form of the momentum equation is:

$$\rho \left[\frac{D\vec{V}}{Dt}\right] = -\nabla p + \nabla \cdot \mu \left[(\nabla \vec{V}) + (\nabla \vec{V})^T\right] + \rho \vec{g} - \gamma \nabla \cdot \left(\frac{\nabla \alpha}{|\nabla \alpha|}\right) \nabla \alpha \tag{14}$$

The discretized form of equations (1), (11), and (14) are solved using the open-source library OpenFOAM. To handle the pressure-velocity coupling, the PIMPLE algorithm [35] is employed in combination with adaptive time stepping, which is based on the CFL (Courant-Friedrichs-Lewy) condition.

## 2.2. Dynamic contact angle model

To accurately predict the shape of the interface in the vicinity of the three-phase contact line, the unit normal to the interface, denoted as $\hat{n}$, is expressed as a function of the dynamic contact angle $\theta_d$ as

$$\hat{n} = \frac{\nabla \gamma}{|\nabla \gamma|} = \hat{n}_w cos\theta_d + \hat{n}_t sin\theta_d \tag{15}$$

Where, $\hat{n}_w$ and $\hat{n}_t$ are the unit vectors of the solid surface in the normal and tangential directions respectively. The contact angle is subject to variation due to existing contact angle hysteresis $(\theta_A - \theta_R)$, leading to a dynamic contact angle that depends on various factors, such as droplet fluid properties, contact line velocity, nature of the substrate, surface interactions of the fluid, and so on. Many well-known contact angle models exist to predict the dynamic contact angle during the spreading or retracting of a drop, such as Shikmurzaev's model [15], Cox model [16], and Kistler's model [36], etc. The contact angle model by Kistler has been incorporated into the VOF solver within the OpenFOAM library, and has been used in this study. Kistler's model is described as

$$\theta_d = f_H \left( Ca + f_H^{-1}(\theta) \right); \tag{16}$$

Where,

$$f_H = COS^{-1}\left(1 - 2tanh\left(5.16\left(\frac{\tau}{1+1.31\tau^{0.99}}\right)^{0.706}\right)\right) \tag{17}$$

The Kistler model (eqn. 16 and 17) describes the dynamic contact angle $(\theta_d)$ manifestation in terms of the contact line velocity $(u_{cl})$ and dynamic wetting process. The $\theta_d$ is expressed as a function of the dynamic contact line Capillary number $\left(Ca = \frac{\mu u_{cl}}{\sigma}\right)$ and the Hoffmann's function $(f_H)$. While implementing the Kistler model, $\theta_d$ is estimated based on the value of $u_{cl}$, as:

$$\theta = \begin{cases} \theta_A, & if\ u_{cl} > 0 \\ \theta_E, & if\ u_{cl} = 0 \\ \theta_R, & if\ u_{cl} < 0 \end{cases} \tag{18}$$

where, $\theta_A, \theta_E\ and\ \theta_R$ represent the advancing contact angle, static equilibrium contact angle, and receding contact angle, respectively. The Kistler model is preferred for its precise prediction of contact line motion. There are certain limitations associated with its use, as it is only applicable within a specific $Ca$ range. Moreover, the bulk motion of the contact line is truly multi-scale. A more accurate prediction of the contact line velocity would require a molecular perspective. Nevertheless, for the length scale of the droplet in the current study, the combination of the Kistler model and the VOF method has been shown to be very accurate in estimating the contact line dynamics [36–38].

### 2.3. Simulation domain and Grid Independence test

A spherical water droplet of post-impact diameter $D_o$ (=2.04 – 3.5 mm) was allowed to impact on the apex of a stationary SH wedge with different impact velocities $U_o$ (refer to Fig. 1 (a)for schematic). The dimensions of 2D simulation domain b×h ($b=6.5D_o$ and $h=6.5D_o$) were carefully chosen to avoid any boundary effects on the hydrodynamics. Both symmetric and asymmetric SH wedges, with different wedge angles $(\phi)$, ranging from 45°-75° have been studied in detail. An inlet-outlet boundary condition is implemented at the top wall of domain to keep it open to the ambient. Zero pressure gradient is imposed on both the side walls of the domain. Additionally, on the surface of the wedge and its associated bottom wall, the Kistler dynamic contact angle model is employed. To impart SH properties to the wedge, the $\theta_A$ and $\theta_R$ are assigned values of 154° and 144° respectively, based on the current experiments. These values aim to minimize the contact angle hysteresis ($CAH = \theta_A - \theta_R = 10°$) in accordance with Kistler's [36] model. These specific values have been obtained from commercial spray-coated (*Neverwet*) SH surfaces. We have studied the role of the droplet diameter in the range $D_o = 2.04 - 3.5$mm.

We have used a structured Cartesian uniform grid to discretize the simulation domain. Various grid sizes were considered in the simulation to obtain the grid-independent results. Fig. 1. (c)illustrates the temporal evolution of the droplet spreading diameter for different grid sizes. Comparing the spreading diameter values between grid sizes of $200 \times 200$ and $250 \times 250$, it was evident that the difference is significantly small; less than 6 μm at equilibrium. The difference between grid sizes of $150 \times 150$ and $200 \times 200$ was higher, approximately 22 μm at equilibrium. Thereby, considering the spreading diameter for both the grid resolutions, $200 \times 200$ grid was selected as the optimal one for the simulations. Additionally, during the lift-off stage from the wedge slopes (refer to Fig. 1. (b)), the shape of the droplet interface profile was tested for different grid sizes. It was observed that the results for both $200 \times 200$ and $250 \times 250$ grid sizes almost overlap perfectly, while a significant deviation was observed between the $150 \times 150$ and $200 \times 200$ (refer to Fig. 1. (b)). Hence, the 200 x 200 grid was justified. Further, we conducted a thorough examination of the temporal spreading diameter variation of droplets on a wedge surface of $\phi_1 = \phi_2 = 45°$ for both 3D and 2D domains. We observed that the maximum deviation between the results is below 5.25% (refer to Fig. 1. (d)). Given the low difference between the 2D and 3D simulation results, we have opted for 2D simulations to reduce computational expenses, wherever applicable without the loss of relevant mechanics. As and where essential in accordance with the nature of the impact or splitting events, we have also performed 3D simulations.

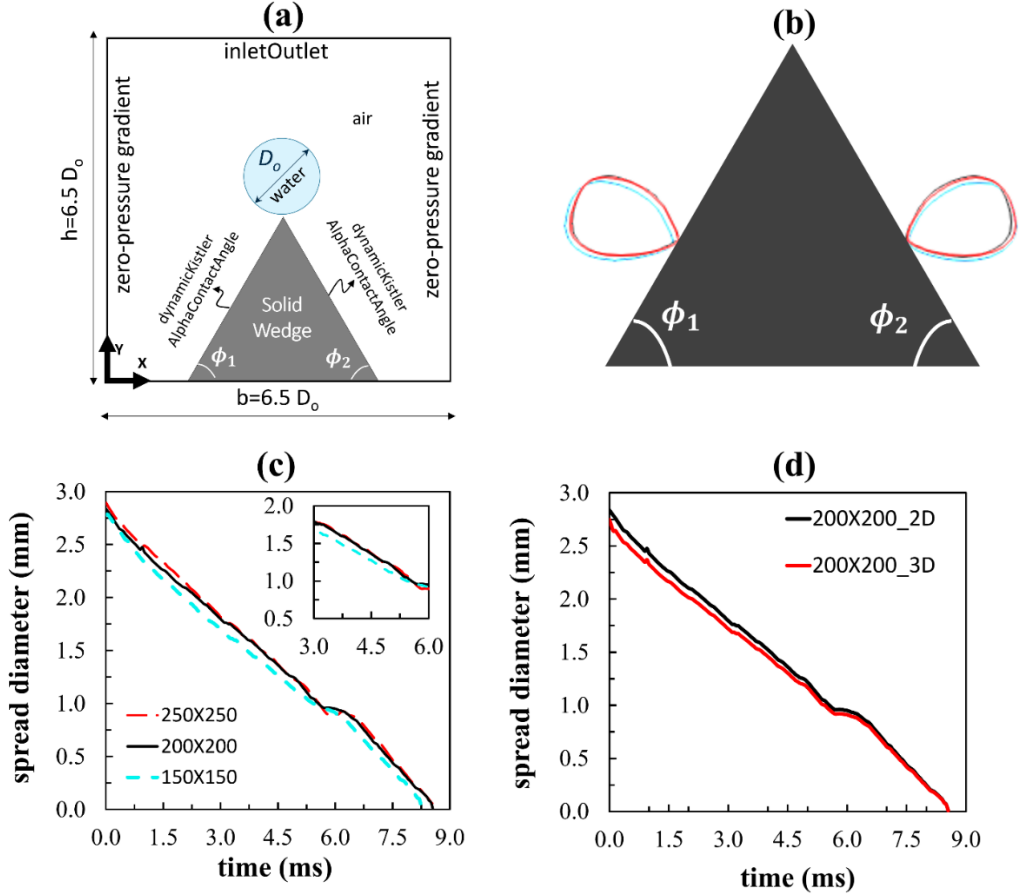

**Fig. 1.** (a) Schematic of the 2D view of the simulation domain of a droplet impact on the apex of a stationary SH wedge. (b) Variation in the shape of the droplet interface profile during daughter droplets lift-off event for different grids as in the legend of part c. (c) 2D grid independence test for different grid sizes for $\phi_1 = \phi_2$. (d) Variation in spreading diameter of the droplet for both 2D and 3D simulations for an optimal grid size of 200x200.

## 3. Experimental methodology

We have performed detailed experiments along the lines of the different simulations. The schematic of the experimental setup is shown in Fig. 2. Here, we used a set of carefully machined and surface-polished aluminum wedges, which were then coated with a commercial superhydrophobic (SH) solution (Neverwet, Ultra Ever dry, USA) [40], [41]. Before coating, each wedge was cleaned with deionized water, followed by acetone, and oven-dried thoroughly to remove the machining oils and debris, dirt and contaminants. The droplets were dispensed on the wedge apex using a digitized droplet dispensing mechanism (DDM) (Holmarc Opto-Mechatronics Ltd., India) integrated with a micro-liter syringe (±0.1μl volumetric accuracy) and a flat head steel needle. The volume of the liquid droplet dispensed was precisely controlled using a digitized controller. To vary the impact Weber number (*We*), the droplets were released from different heights, realized using a motorized x-y-z stage attached to the DDM. All hydrodynamic events were recorded using a high-speed camera (Photron, UK) equipped with a 105 mm focal length macro lens. All the events were recorded in shadowgraph mode at 4000 frames per second, with a resolution of 1024×1024, and shutter speed of 2.7 µs.

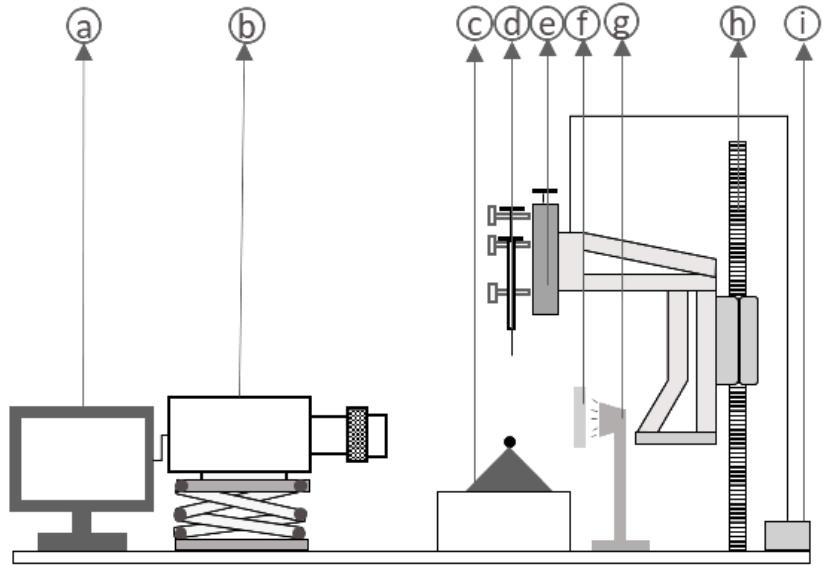

**Fig. 2.** Schematic of the experimental setup: (a) computer system for camera control and data acquisition (b) high-speed camera (c) superhydrophobic (SH) wedge target (d) microliter syringe (e) droplet dispensing mechanism (DDM) unit (f) diffuser (g) strobe light (h) motorized x-y-z stage (i) DDM controller

## 4. Results and discussions

### 4.1. Role of wedge angle ($\phi$) on spreading dynamics for symmetric wedges

First, we discuss the case of symmetric wedges. We have explored different wedge angles, ranging between $\phi = 45^o\text{-}75^o$, while maintaining a constant $We=9.25$. Fig. 3. (a-c) illustrate the role of the $\phi$ on the temporal evolution of the droplet spreading, starting from the apex of the wedge. The spread factor ($\beta=D/D_o$) is defined as a ratio of spreading diameter($D$) at any instant to pre-impact droplet ($D_o$). From Fig. 3. (a) and (b), it is evident that the $\beta$ decreases as $\phi$ increases due to the sharper inclination of the wedge. A sharper wedge apex leads to diminished available contact area, resulting in a reduced extent of droplet spreading across the surface. Our experiments corroborate our simulations (see Fig. 3. (c)), demonstrating a significant relationship between $\beta$ and the non-dimensional time ($\tau = tUo/D_o$), where t, $U_o$, and $D_o$ represent the elapsed time, impact velocity, and pre-impact droplet diameter, respectively. An average deviation of only ~8% is noted between the experiments and 2D simulations in Fig. 3. It is important to note that the splitting of the droplet at the wedge apex is accurately predicted by the simulations. Also, during the lift-off away from the sides of the wedge (see 6$^{th}$ column of Fig. 3.

(a)), a major observation is that the droplet's lift-off occurs relatively faster than anticipated by the simulation results, irrespective of $\phi$. This mismatch could arise due to the 2D nature of these simulations.

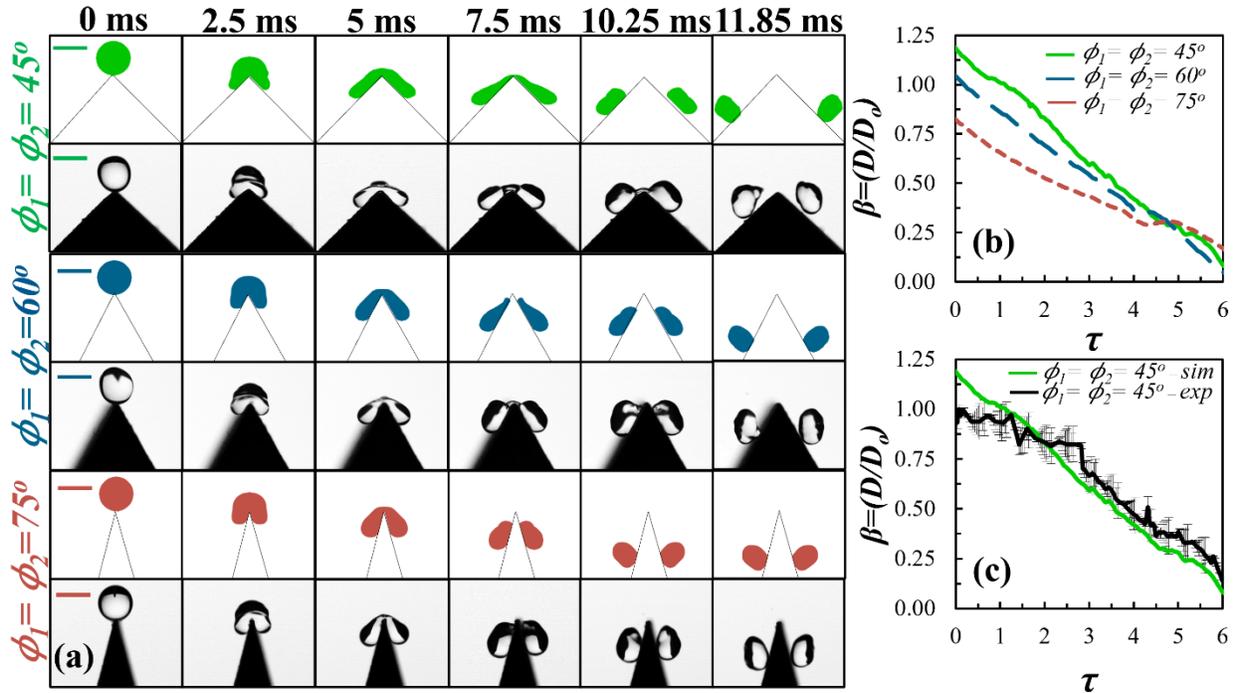

**Fig. 3.** Effect of the wedge angle *(ϕ)* on the spreading dynamics: (a) Temporal evolutions of the droplet for a fixed *We*~9.25 (each odd and even row represents simulation and corresponding experimental observations, respectively) (b)variation in *β* for different *ϕ* from simulation (c) experimental comparison of the *β* for $\phi_1 = \phi_2 = 45^o$. The scale bars in Fig. 3. (a) represent 2.8 mm.

### 4.2. Role of *We* on maximum spread factor *(β)$_{max}$* of the droplet for symmetric wedges

Next, we examine the effect of *We* on the maximum spread factor *(β)$_{max}$*=$D_{max}/D_o$ of the droplet on symmetric wedges. As observed from Fig. 4. (a-b), it is evident that at a particular *ϕ*, the *(β)$_{max}$* increases with an increase in the *We;* due to the higher inertial forces during impact. It is also seen that the *(β)$_{max}$* decreases with an increase in *ϕ* at a particular *We;* due to the sharp steepness of the wedge, which provides less contact area for the droplet to spread out. We have also verified the simulation results with our experiments and found that the highest *(β)$_{max}$* on a wedge of $\phi_1 = \phi_2 = 60^o$, over various *We* agree quite closely, with a maximum of ~9% deviation.

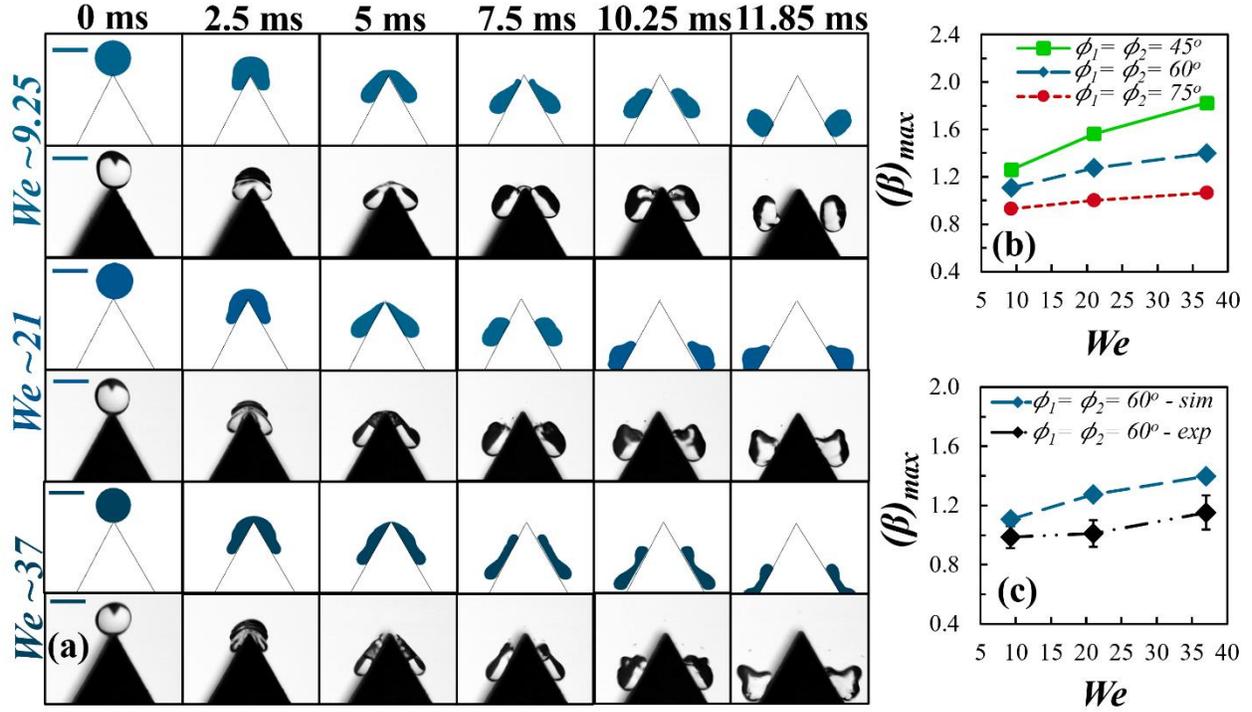

**Fig. 4.** Effect of Weber number (*We)* on the spreading dynamics: (a) temporal evolutions of the droplet for $\phi_1 = \phi_2 = 60^o$ (each odd and even row represents simulation and corresponding experimental observations, respectively) (b) variation in *(β)$_{max}$* for different $\phi$ from simulations (c) experimental comparison of the *(β)$_{max}$* for $\phi_1 = \phi_2 = 60^o$. The scale bars in Fig. 4. (a) represent 2.8mm.

### 4.3. Role of wedge asymmetry $(\phi_1 - \phi_2)$ on the spreading dynamics for asymmetric wedges

Here, we probe the role of the wedge asymmetry *($\phi_1 45^o$- $\phi_2 60^o$)* on spreading dynamics at a fixed *We*~9.25. As shown in Fig. 5. (a-b), on asymmetric wedges the *(β)* increases relatively more along the less steep side *($\phi_1 = 45^o$)* than on the steeper side *($\phi_2 = 60^o$)*. This is because one side *($\phi_1 = 45^o$)* of the wedge allows spreading more easily due to a larger contact area compared to the other side *($\phi_2 = 60^o$)*. The observations are likewise on the *($\phi_1 75^o$- $\phi_2 60^o$)* asymmetric wedge. We have also verified the behavior of *β* with *τ* for the *($\phi_1 45^o$- $\phi_2 60^o$)* wedge from the experiments (see Fig.5. (c)). Good agreement with the simulations is noted, albeit with a maximum deviation of ~11%. It is also important to mention here that while the splitting event of the droplet is accurately predicted by the 2D simulations, the lift-off after the split is not captured accurately (last column, Fig. 5. (a)). This is especially true for the strongly asymmetric wedge *($\phi_1 75^o$- $\phi_2 60^o$)*. One potential reason for the mismatch in the lift-off event for asymmetric wedges (last column, Fig. 5(a)) may be the non-uniformity of the superhydrophobic coating on the wedge surface. Another reason could be the asymmetry of the wedge surface, which affects the timescales of the parent droplet's spreading and the daughter droplets' retraction before lift-off on both sides. To investigate these possible explanations in more detail, we extended our analysis through impact force tracing in the following section 4.9.

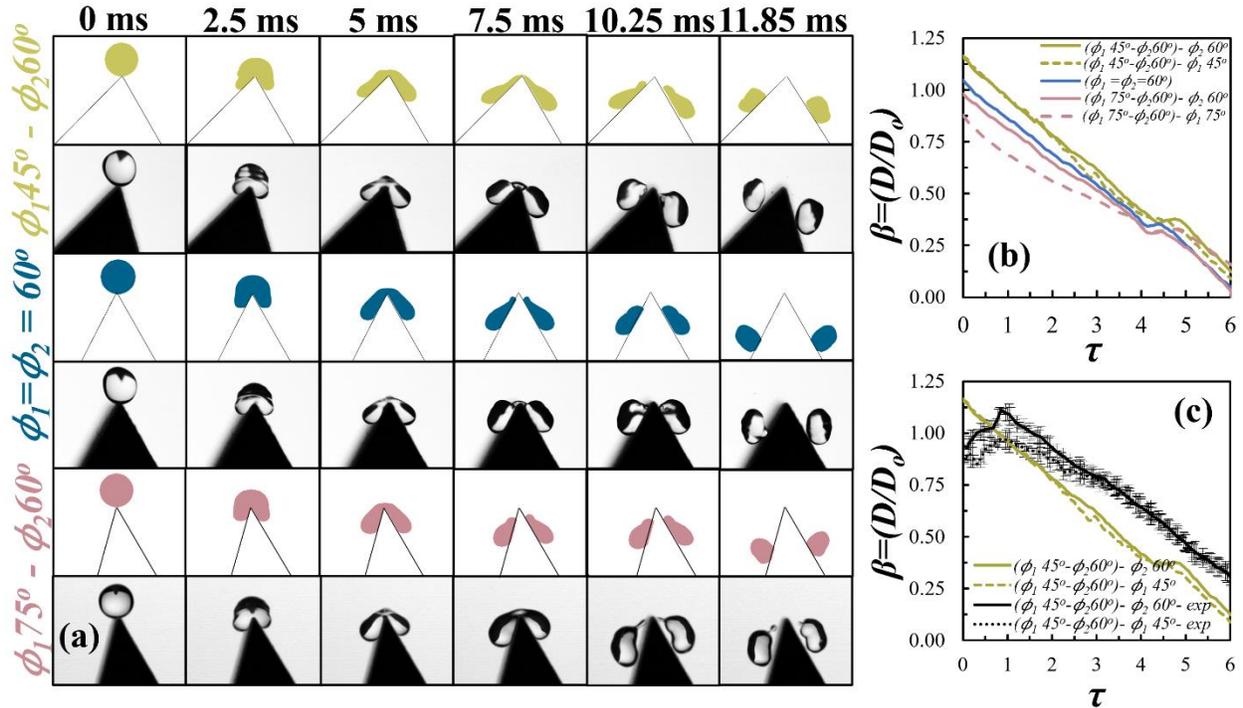

**Fig. 5.** Influence of the wedge asymmetry angle $(\phi_1 - \phi_2)$ on the spreading dynamics: (a) temporal evolutions for a fixed *We* ~ 9.25 (each odd and even row represents simulation and corresponding experimental observations, respectively) (b) variation in *β* for different $(\phi_1 - \phi_2)$ (c) experimental comparison of *β* for *($\phi_1 45^o$- $\phi_2 60^o$)*. The scale bars in Fig. 5. (a) represent 2.8mm.

### 4.4. Role of *We* on *(β)$_{max}$* for asymmetric wedges

As shown in Fig. 6. (a-b), the *(β)$_{max}$* increases with an increase in the *We* due to the higher inertial forces available during impact at the apex. Also, it is observed that the less steeper side *(corresponding to a smaller inclination angle)* results in a relatively larger *(β)$_{max}$* compared to the steeper side (with a larger inclination angle). This is attributed to the greater available contact area for the droplet on the gentler slope. A similar behavior is also observed in *($\phi_1 75^o$- $\phi_2 60^o$)*. We have compared our simulation outcomes with experimental data, and a deviation of only ~6% is observed.

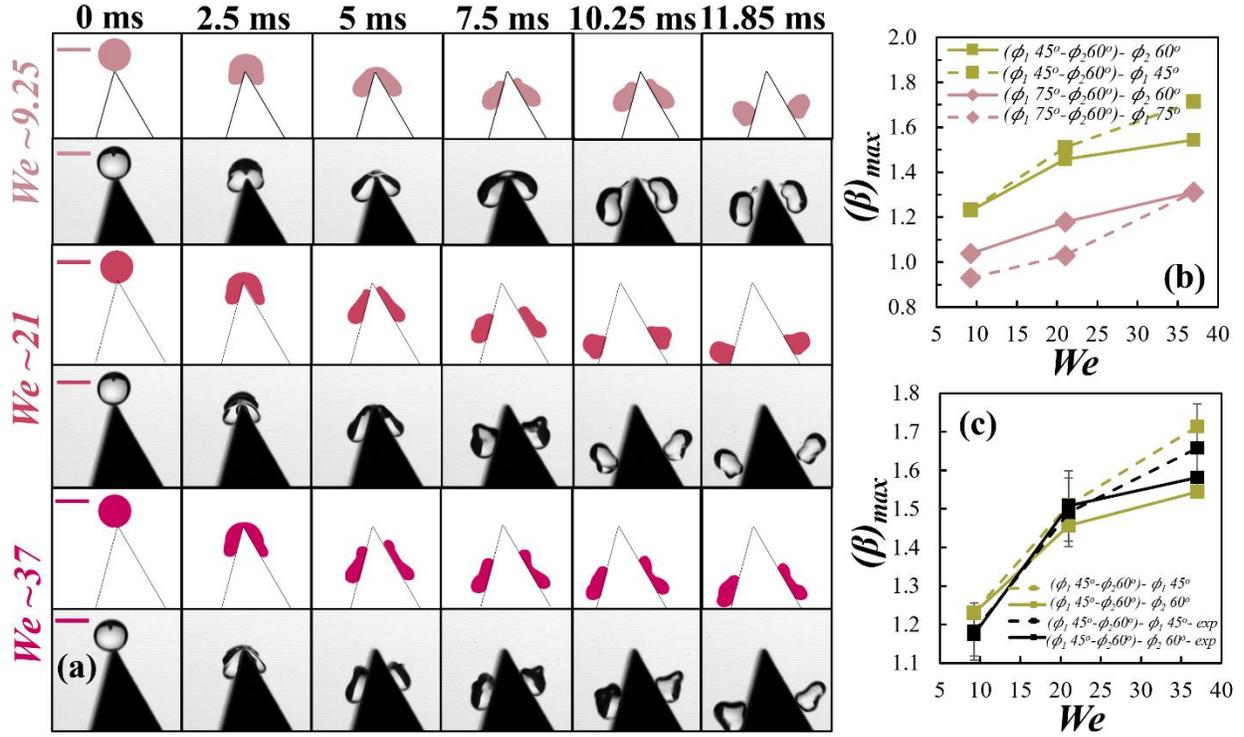

**Fig. 6.** Influence of the *We* on the spreading dynamics: (a) Temporal evolutions of the droplet for $(\phi_1 75^o\text{-}\phi_2 60^o)$ (each odd and even row represents simulation and corresponding experimental observations, respectively) (b) maximum spread factor $(\beta)_{max}=D_{max}/D_o$ for different wedge angles (c) experimental comparison of the $(\beta)_{max}$ for $(\phi_1 45^o\text{-}\phi_2 60^o)$. The scale bars in Fig. 6. (a) represent 2.8mm.

### 4.5. Role of normalized Bond number ($Bo^*$) on the spreading dynamics for symmetric wedges

We now discuss the effect of the normalized Bond number $(Bo^* = Bo/(Bo)_{capillary})$ on the spreading dynamics for symmetric wedges. We have defined the $(Bo^*)$ as the ratio of the operating Bond number $(Bo)$ (which is the ratio of gravitational force to the surface tension force, $(Bo=\rho g D_o^2/\sigma)$, where $\rho, g, D_o$ and $\sigma$ are density, acceleration due to gravity, pre-impact droplet diameter and surface tension respectively.) to the Bond number for the capillary length scale $((Bo)_{capillary} = \rho g D_o^2/\sigma$, where $D_o$ =2.8mm in all operating cases) of the fluid (water here). We have varied the $D_o$ ranging from 2.04 – 3.5 mm to vary the $Bo$.

Fig.7. (a). shows the temporal evolution of the droplet at different $Bo^*$ on a symmetric wedge $(\phi_1 = \phi_2 = 45^o)$. Droplets with higher $Bo^*$ (= 1.5) lift off the wedge surface quicker than those with low $Bo^*$ (= 0.5) due to the dominance of gravitational force over capillary force. Although theoretically counterintuitive, as stronger gravitational forces should cause more spreading, we observed that larger $Bo^*$ droplets exhibit earlier lift-off with intermediate liquid bridge formation (refer to 6$^{th}$ row, 5$^{th}$ column of Fig. 7(a)). This may result from the inertial collapse of the droplet at the onset of splitting, leading to immediate lift-off with intermediate liquid bridge formation. Fig.7. (b). shows that $\beta$ decreases with increasing $\tau$ for a fixed $Bo^*$, and the trend shifts upward with larger $D_o$.

Fig.7. (c). illustrates that $(\beta)_{max}$ increases with $Bo^*$ at a given $We$ due to dominant gravitational forces and also increases with $We$ at a fixed $Bo^*$ due to higher inertial forces. Fig.7. (d). shows that $(\beta)_{max}$ decreases with increasing $\phi$ for a given $Bo^*$ and $We$ due to the wedge's steepness, which reduces the contact area for spreading. Furthermore, it is also evident that at fixed $\phi$ and $We$, $(\beta)_{max}$ increases with $Bo^*$ due to dominant gravitational forces over capillary forces.

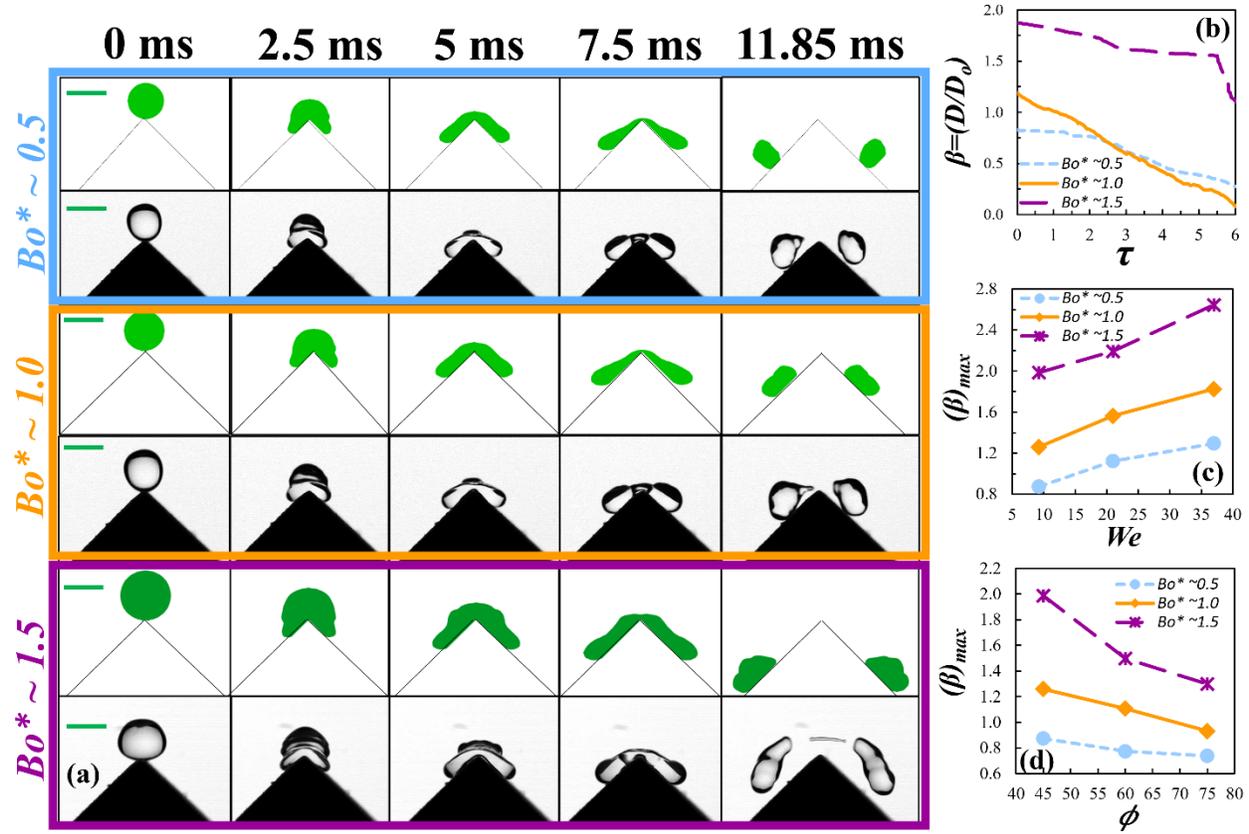

**Fig. 7.** Effect of the $Bo^*$ on the spreading dynamics at a fixed $We$~9.25 for $(\phi_1 = \phi_2 = 45^o)$: (a) snapshots of the time evolutions of droplet over different $Bo^*$ (each odd and even row represents simulation and corresponding experimental observations respectively) (b) variation in $\beta$ under varying $Bo^*$ (c) change in $(\beta)_{max}$ over different $Bo^*$ against various $We$ (d) variation in $(\beta)_{max}$ for different $\phi$. The scale bars in Fig.7. (a) represent 2.8mm.

### 4.6. Role of $Bo^*$ on the spreading dynamics for asymmetric wedges

Fig. 8. (a). presents the temporal evolutions of the droplet on an asymmetric wedge($\phi_1 75^o$-$\phi_2 60^o$). From Fig. 8. (b), it is evident that the trend of the $\beta$ over $\tau$ shifts upwards with an increase in $Bo^*$ at a particular $We$~9.25. This is due to the dominance of gravitational forces against capillary forces. Besides, for a fixed $Bo^*$ and $We$, the less steeper side ($\phi_2 = 60^o$) of the wedge allows for easier droplet spreading due to greater contact area compared to the steeper side ($\phi_1 = 75^o$).

Similarly, Fig. 8. (c). shows the change in $(\beta)_{max}$ across different $Bo^*$ at varying $We$, It is noted that the $(\beta)_{max}$ increases with an increase in $Bo^*$ at a fixed $We$ due to the dominant gravitational forces than capillary forces. Further, it is also noticed that at the same $Bo^*$, the $(\beta)_{max}$ increases with an increase in $We$ due to the dominant inertial forces than capillary forces. Among all operated cases, the less steep and large $Bo^*$ at higher $We$ exhibit the larger $(\beta)_{max}$ of the droplet due to the dominance of both inertial and gravitational forces over the capillary forces. In contrast to the symmetric case ($\phi_1 = \phi_2 = 45^o$), the asymmetric case ($\phi_1 75^o$- $\phi_2 60^o$) shows earlier droplet lift-off at low $Bo^*$ and fixed $We$. This is due to the strong influence of wedge asymmetry. Since both sides of the asymmetric wedge have unequal spreading areas and lift-off time scales, on both sides hinder immediate lift-off during the split-up phase, thus suppressing post-split lift-off.

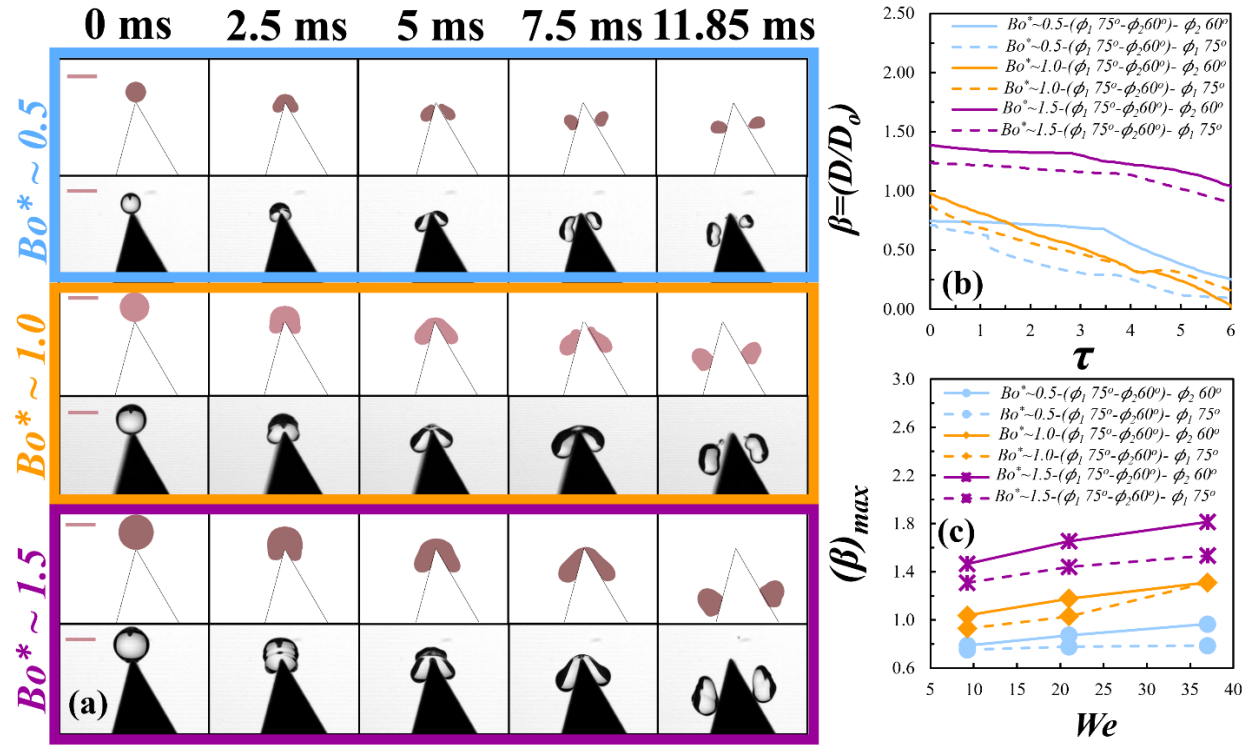

**Fig. 8.** Effect of the $Bo^*$ on the spreading dynamics for ($\phi_1 75^o$- $\phi_2 60^o$) asymmetric wedge: (a) temporal evolutions of droplet for different $Bo^*$ (each odd and even rows represents simulation and corresponding experimental observations respectively) (b) $\beta$ over varying $Bo^*$ at a fixed

*We*~9.25 (c) change in *(β)max* for different *Bo*$^*$ against various *We*. The scale bars in Fig.8. (a) represent 2.8mm.

### 4.7. Role of wedge asymmetry on the split volume of the droplet

Through 3D numerical simulations, the variation in the split volume of daughter droplets during the split-up phase over different $\phi$ has been depicted in Fig. 9. (a). It shows an increase in split volume with increasing $\phi$ from 45° to 75° at a fixed *We*~9.25, due to increased wedge steepness promoting more significant droplet splitting. Similarly, we show the variation in split volumes of the daughter droplets during the split-up phase on asymmetric wedges *($\phi_1 - \phi_2$)* at the same *We* ~9.25 in Fig. 9. (b). It reveals that the gentler side *($\phi_1 45°$)* of the wedge generates a smaller volume of the daughter droplet compared to the steeper side *( $\phi_2 75°$)*.

Here it has been noted that the less steep side $\phi_1 = 45°$ (refer to blue color bar in Fig. 9. (b)) of *($\phi_1 45°$- $\phi_2 60°$)* wedge generates a smaller volume of the daughter droplet than the steeper side $\phi_2 = 60°$ of the same asymmetric wedge. Likewise, *($\phi_1 75°$- $\phi_2 60°$)* generates more volume of the daughter droplets on the steeper side $\phi_1 = 75°$ (refer to red color bar in Fig. 9. (b).) than the gentler side *($\phi_2 = 60°$)* due to the lesser spreading area available for the droplet to spread out against the splitting event.

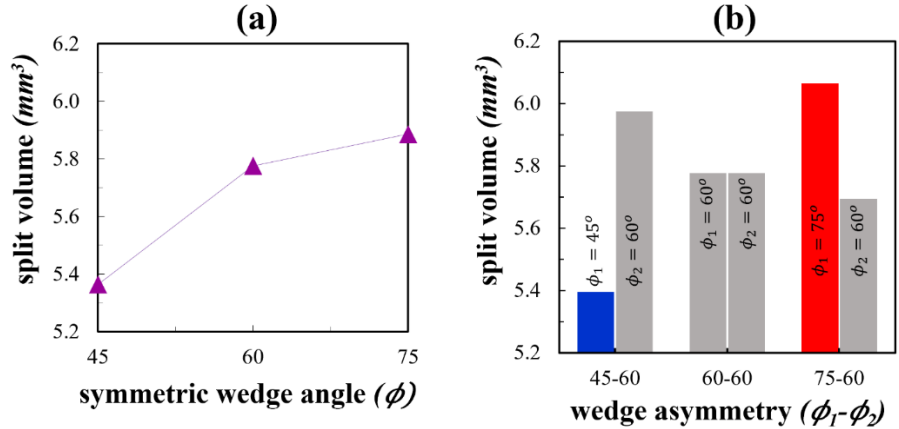

**Fig. 9.** Numerical 3D simulation results of variation in split volume of daughter droplets during split-stage of the parent droplet(a) for different $\phi$ (b) for different *($\phi_1 - \phi_2$)*

### 4.8. Energy budget of the post-impact droplet on symmetric wedges

We observed a certain degree of deviation (in the tune of ~5– 11 %) of the 2D simulation results with our experimental observations in sub-sections 4.1-4.6. More importantly, the 2D model could not accurately predict the late-stage or splitting-hydrodynamics of the droplets. Here, we extend the analysis to the 3D simulations to understand the energy budget of the post-impact droplets. Also, our objective remains to see if the 3D simulations can accurately track the splitting hydrodynamics. Unlike on horizontal SH surfaces, we have noticed the droplet to slide along the slant edges of the SH wedge, along with droplet split-up, retraction and lift-off events. We analyze droplet shape deformation and energy conversion in a complete cycle from pre-impact to lift-off of the daughter droplets from the wedge surfaces[20], [42].

The three relevant energy components for droplet impact are: kinetic energy ($E_K^o$), potential energy($E_P^o$) and surface energy ($E_S^o$). The total initial energy of the droplet just prior to the impact is $E_0 = E_K^o + E_P^o + E_S^o$. Energy conservation at time *t* post-impact yields:

$$E_0 = E_P\,^{(t)} + E_K\,^{(t)} + E_D\,^{(t)} + E_S\,^{(t)} \tag{19}$$

where, $E_0$, $E_P\,^{(t)}$, $E_K\,^{(t)}$, $E_D\,^{(t)}$, $E_S\,^{(t)}$ represent the total initial energy, potential energy, kinetic energy, viscous dissipation energy and surface energy at time *t*. Further, to account for energy loss to the surrounding air, kinetic energies are separately considered for the droplet $E_{K,d}\,^{(t)}$, and ambient gas (air) $E_{K,g}\,^{(t)}$. The energy components are determined by evaluating integrals from the numerical simulations as:

$$E_{K,d}\,^{(t)} = \tfrac{1}{2}\int_{V_d}\rho_d|V|^2 dV_d \tag{20}$$

$$E_{K,g}\,^{(t)} = \tfrac{1}{2}\int_{V_g}\rho_g|V|^2 dV_g \tag{21}$$

$$E_P\,^{(t)} = \rho_d g \int_{V_d} h_c dV_d \tag{22}$$

The viscous dissipation within the droplet is obtained using equations 23-25, as:

$$E_D\,^{(t)} = \int_0^t W(t)dt \tag{23}$$

Here, the viscous dissipation rate $W(t)$ is evaluated in terms of the stress tensor **T** and the rate of the stress tensor **S**, as:

$$W(t) = \int_{V_d \cup V_g} \mathbf{T}:\mathbf{S}\,dV \tag{24}$$

$$W(t) = \int_{V_d \cup V_g}\left[2\mu\left[\left(\tfrac{\partial V_x}{\partial x}\right)^2 + \left(\tfrac{\partial V_y}{\partial y}\right)^2 + \left(\tfrac{\partial V_z}{\partial z}\right)^2 - \tfrac{1}{3}(\nabla \cdot V)^2\right] + \mu\left[\left(\tfrac{\partial V_y}{\partial x} + \tfrac{\partial V_x}{\partial y}\right)^2 + \left(\tfrac{\partial V_z}{\partial y} + \tfrac{\partial V_y}{\partial z}\right)^2 + \left(\tfrac{\partial V_x}{\partial z} + \tfrac{\partial V_z}{\partial x}\right)^2\right]\right]dV \tag{25}$$

The surface energy of the droplet is calculated from:

$$E_S\,^{(t)} = \int_A \gamma dA \tag{26}$$

where *A* represents the surface area of the droplet.

In this context, it is noticed that when a droplet impacts the apex of a ($\phi_1 = \phi_2 = 45°$) wedge with *We*~9.25 (refer to Fig. 10. (a)), it spreads along the slant edge without splitting until 5 ms (refer to 3rd column of Fig. 10 (a-d)). At ~7.5 ms, it splits into two nearly equal daughter droplets, which spread and slide along the edge, lifting off at ~11.85 ms (refer to 5$^{th}$ column of Fig. 10. (a-d)). After the drop impact on the wedge, the kinetic energy of the droplet decreases while surface energy increases during spreading. Post-split, the daughter droplets continue spreading and eventually reaches maximum surface energy at maximum spread state. Subsequently, the daughter droplets retract with increased kinetic energy until lift-off. During the lift-off, viscous dissipation energy rises as the droplets become ellipsoidal (refer to Fig.10. (d) and red diamond trend line in Fig. 10. (e)). As shown in Fig. 10. (f), increase of the wedge angle from $\phi = 45^o$ to $\phi = 75^o$ reduces the split time from ~6.25 ms to ~5 ms due to the variation in the steepness. Velocity profiles (refer to Fig.10. (f)) show that higher $\phi$ cause the droplet to split and roll post-impact, due to reduced spread area. Fig. 10. (g) exhibits pressure peaks during pre-impact and split-up stages, caused by sudden changes in three-phase contact line velocities and high capillary wave oscillations[43] at the liquid-air interface during spreading. We believe the significant changes in velocity contours and pressure distribution are due to the triple contact line on the wedge surface acting as a moving wave source [43], generating local curvature gradients. This causes capillary wave oscillations at the onset of split-up, leading daughter droplets to roll instead of slide (refer to Fig. 10. (f)., 4$^{th}$ and 5$^{th}$ columns).

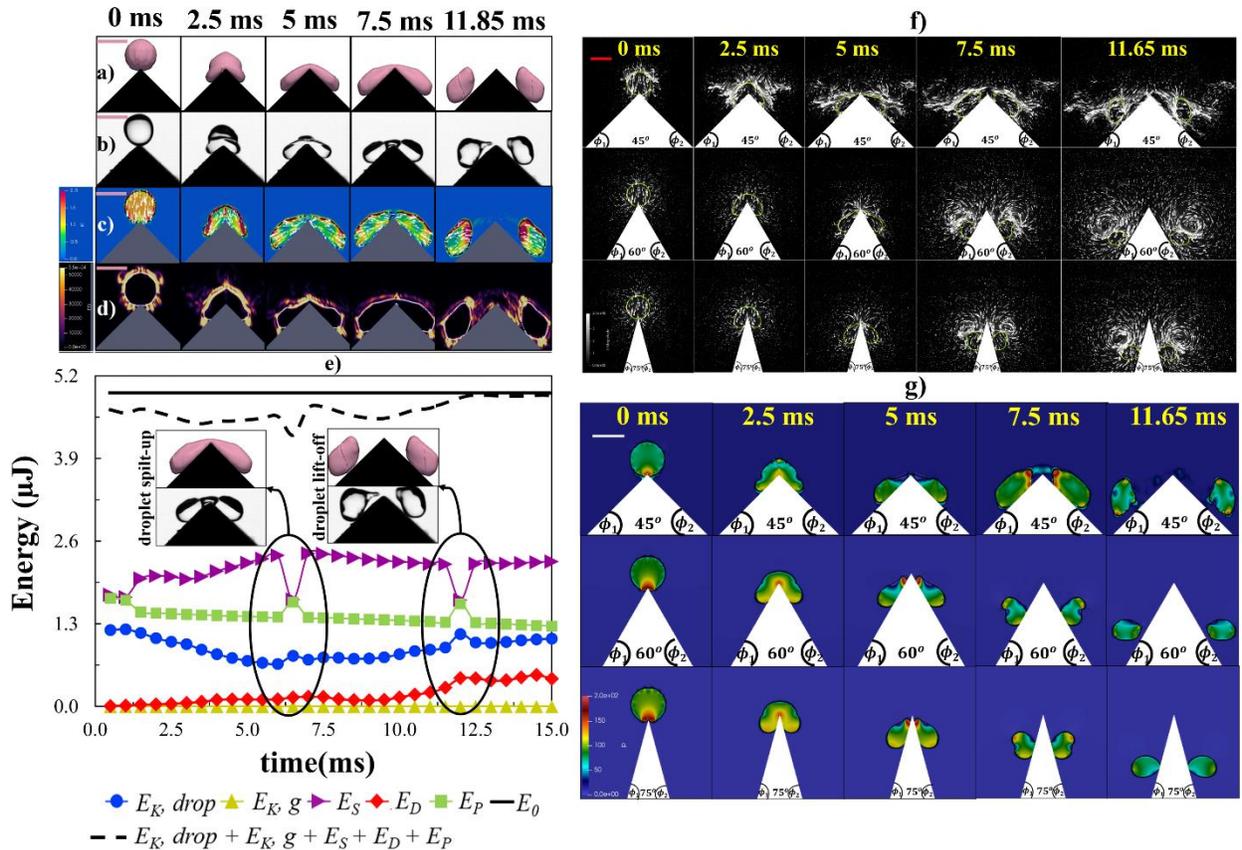

**Fig. 10.** Temporal evolutions of the 3D droplet for ($\phi_1 = \phi_2 = 45^o$) at a fixed *We*~9.25 (a) 3D simulation (b) Experimental (c) Kinetic energy distribution across the droplet per unit volume (µJ/m$^3$) (d)Viscous dissipation energy distribution across the droplet per unit volume per unit

time (µJ/m³ s) and (e) Energy budget estimation. (f) Velocity distribution (g) Pressure variation. The scale bars in Fig.10. (a-g) represent 2.8mm.

## 4.9. Impact force history

Finally, through numerical simulations, we tried to trace the history of impact force $F(t)$ of a 3D droplet on both symmetric $(\phi_1 = \phi_2 = 45^o)$ and asymmetric $(\phi_1 45^o\text{-}\phi_2 60^o)$ SH wedge surfaces at $We \sim 9.25$ (refer to Fig. 11.).The droplet impact force on the surface is obtained by integrating the pressure field on the wedge as shown in equation (27).

$$F(t) = \int_A (p) dA \qquad (27)$$

On one hand, when a droplet impacts on the apex portion *(t~0ms)* of the symmetric $(\phi_1 = \phi_2 = 45^o)$ SH wedge, there is a sudden rise in its impact force and reaches a maximum value of $F_{1_{(\phi_1=45^o)}} = F_{1_{(\phi_2=45^o)}} = 0.0045 \mu N$ (refer to the state B in Fig. 11.) during the onset of the spreading stage *(t~2.6ms)*. After that, the parent droplet starts to spread and reaches a minimum value of $0.0038\mu N$ at the maximum spreading stage without split *(t~5ms)*. The force continues to decrease to $0.00286$ µN as the parent droplet splits into two nearly equal daughter droplets *(t~7.25ms)*. Once the droplet splits up into two daughter droplets, the two daughter droplets separately retract against sliding on two slant symmetric SH wedge sides. Due to the large amount of local viscous dissipation and capillary waves [43] generation at the interface of the droplet, its shape gets transformed from spheroidal to ellipsoid. At this instance, we have noted that there is a gradual secondary peak in the impact force (refer to the state E in Fig. 11.) of $F_{2_{(\phi_1=45^o)}} = F_{2_{(\phi_2=45^o)}} = 0.00318 \mu N$ during the onset of retraction stages of two daughter droplets against sliding at *t~8.5ms* due to the stored impact kinetic energy. Subsequently, following this secondary peak (refer to state E in Fig. 11), the impact force steadily decreases until it reaches zero as the two daughter droplets lift off *(t~10.75ms)* the wedge sides.

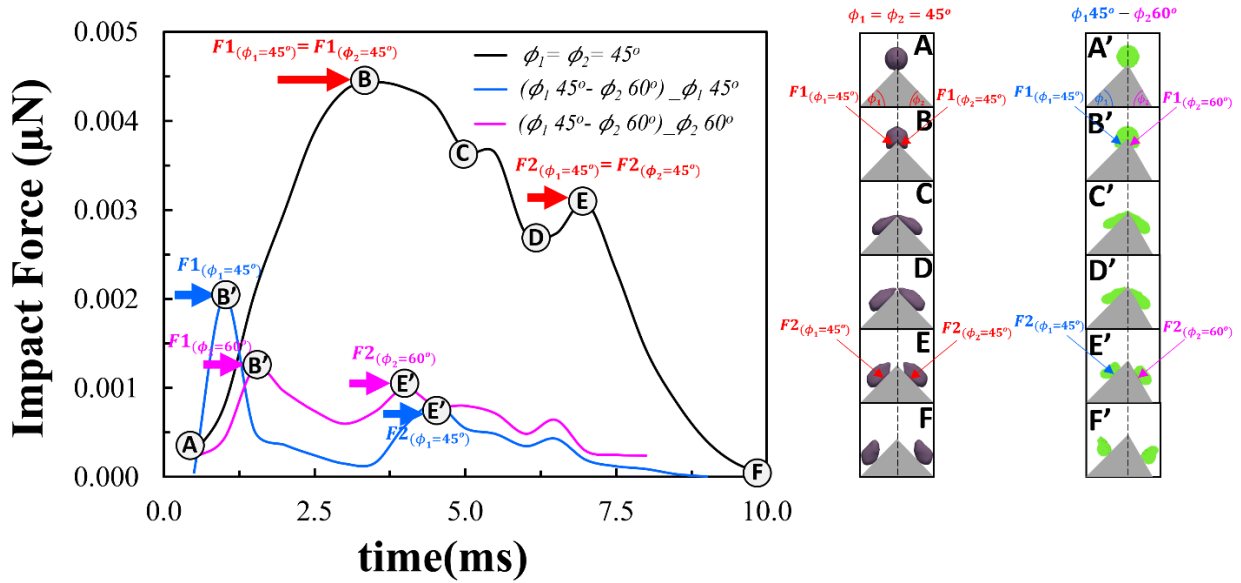

**Fig. 11.** Impact force history of the 3D-droplet on $(\phi_1 = \phi_2 = 45^o)$ and $(\phi_1 45^o\text{-}\phi_2 60^o)$ wedge surfaces at $We\sim9.25$.

On the other hand, when the same droplet impacts the apex portion $(t\sim0ms)$ of an asymmetric SH wedge $(\phi_1 45^o\text{-}\phi_2 60^o)$, as similar to symmetric $(\phi_1 = \phi_2 = 45^o)$ case, initially there is a sudden rise in its impact force and reaches a peak $F_{1_{(\phi_1=45^o)}} = 0.0021\mu N$ and $F_{1_{(\phi_2=60^o)}} = 0.00145\mu N$ on the less steeper $(\phi_1 = 45^o)$ and steeper $(\phi_2 = 60^o)$ sides respectively (refer to B' on both blue and pink color trend lines in Fig. 11.). It is due to the change in unequal available spreading areas, which creates strong competition between spreading and splitting. Consequently, there is a mismatch in the time intervals for the droplet to split on both sides of the wedge. Similarly, post split-up, the retraction and lift-off time scales of daughter droplets are also different due to the unequal volumes of the daughter droplets generation in the asymmetric wedge case. It led to an abrupt change in the secondary peak in impact force $F_{2_{(\phi_1=45^o)}} = 0.00085\mu N$ and $F_{2_{(\phi_2=60^o)}} = 0.00105\mu N$ on the gentler $(\phi_1 = 45^o)$ and steeper $(\phi_2 = 60^o)$ sides respectively (refer to E' on both blue and pink color trend lines in Fig. 11.).

## 5. Conclusions

In summary, for symmetric SH wedges, we found that water droplets spread more easily as $\phi$ decreases rather than rolling, even with nearly equal volumes of the daughter droplets at a fixed *We*. Droplets exhibit larger spreading with an increase in *We* for a fixed wedge angle $(\phi)$ due to the dominant inertial forces than capillary forces. It also shows that $(\beta)_{max}$ decreases with an increase in $\phi$ at a fixed *We* due to the greater steepness, which reduces the available contact area for droplet spreading. For asymmetric wedges $(\phi_1 - \phi_2)$, we found that the spread factor $(\beta)$ increases relatively more on the gentler side than the steeper side. Post split, the parent droplet generates unequal volumes of daughter droplets under varying *We*. It is revealed that, among all operated cases, $(\phi_1 75^o\text{-}\phi_2 60^o)$ splits a larger volume of the daughter droplets on the steeper side $(\phi_1 75^o)$ due to the less spread area of the droplet to spread out against splitting. For both symmetric and asymmetric cases, the $(\beta)_{max}$ of the droplet increases with an increase in $Bo^*$ due to a larger $D_o$ regardless of at a particular *We*. By estimating the energy budget of the complete lift-off cycle of an impacting water droplet on a symmetric SH wedge, it is observed that there is a strong mutual competition between surface energy and stored kinetic energy during the retraction phase of the droplet, which eventually leads to become a factor for the significant rise in its viscous dissipation in the split-up to lift-off stages (refer to 4-6$^{th}$ columns of Fig.10(c-d)). Our numerical evaluation of the impact force history (refer to Fig. 11.) revealed significant fluctuations with peaks during the onset of droplet spreading and daughter droplet retraction phases on both symmetric and asymmetric SH wedges. Our findings could be useful for estimating and understand the intricacies of droplet splitting against stationary or moving obstacles in micro- and macro-fluidic scenarios[44].

**Data availability:** All data pertaining to this research work is available in the manuscript.

**Conflicts of interests:** The authors do not have any conflicts of interest with respect to the current research work.

**Author Contributions: Gudlavalleti V V S Vara Prasad:** Conceptualization; Validation; Data curation; Formal analysis; Investigation; Methodology; Writing – original draft. **Parmod Kumar:** Conceptualization; Formal analysis; Investigation; Methodology; Supervision; Writing – review & editing. **Purbarun Dhar:** Conceptualization; Formal analysis; Investigation; Methodology; Supervision; Writing – original draft. **Devranjan Samanta:** Conceptualization; Validation; Formal analysis; Investigation; Methodology; Supervision; Writing – review & editing.

**Acknowledgments:** GVVSVP would like to thank the Ministry of Education, Govt. of India, for the doctoral scholarship. GVVSVP would like to thank Debarshi Debnath for his helpful support and fruitful discussions for the present work. PK would like to thank IIT Mandi for providing computational resources for the present work. All authors would like to acknowledge National Supercomputing Mission (NSM) for providing computing resources of 'PARAM Himalaya at IIT Mandi, which is implemented by C-DAC and supported by the Ministry of Electronics and Information Technology (MeitY) and Department of Science and Technology (DST), Government of India.